%% file: ITW_Submission.tex
\documentclass[conference,letterpaper]{IEEEtran}
\IEEEoverridecommandlockouts

\def\Figs{./figs/} 

\usepackage[utf8]{inputenc} 
\usepackage[T1]{fontenc}
\usepackage{url}
\usepackage{ifthen} 
\newboolean{short_version}
\setboolean{short_version}{false} 
\usepackage[cmex10]{amsmath} 

\usepackage[notheorems,IEEEtran]{research17} 
\usepackage[
  top=0.700in, 
  bottom=1.080in,
  left=0.673in,
  right=0.673in
]{geometry}
\usepackage{balance} 
\usepackage{amssymb,amsfonts,amsbsy}
\usepackage{enumitem}
\usepackage{mathtools}
\usepackage[lined,boxed,commentsnumbered,linesnumbered,ruled]{algorithm2e}
\usepackage{comment}
\setlist[description]{leftmargin=\parindent,labelindent=\parindent}

\newtheorem{theorem}{\mytheoremname}
\newtheorem{lemma}{\mylemmaname}

\newtheorem{conjecture}{\myconjecturename}
\newtheorem{proposition}{\mypropositionname}

\newtheorem{definition}{\mydefinitionname}
\newtheorem{remark}{\myremarkname}
\newtheorem{example}{\myexamplename}
\newcommand{\Hwt}[1]{\wH\left(#1\right)} 
\newcommand{\ConstrAq}[2][]{\Lambda_{\textnormal{A}_{#1}}(#2)} 

\newcommand*{\Scale}[2][4]{\scalebox{#1}{\ensuremath{#2}}} 
\makeatletter
\renewcommand*\env@matrix[1][*\c@MaxMatrixCols c]{%
  \hskip -\arraycolsep
  \let\@ifnextchar\new@ifnextchar
  \array{#1}}
\makeatother

\usepackage{tikz}
\usetikzlibrary{calc, shapes, patterns, decorations.text, decorations.pathreplacing}
\usetikzlibrary{spy} 
\usepackage{ctable} 
\usepackage{afterpage} 
\usepackage{lscape} 
\usepackage{pdflscape} 
\usepackage{rotating} 
\usepackage{pbox} 
\usepackage{nicematrix}
\usepackage{centernot}
\usepackage{stmaryrd}
\usepackage{soul}
\allowdisplaybreaks

\definecolor{lightorange}{rgb}{1.0, 0.63, 0.48}
\definecolor{celadon}{rgb}{0.67, 0.88, 0.69}

\usepackage{soul} 
\usepackage[colorinlistoftodos, textsize=footnotesize]{todonotes}

\begin{document}
\title{Generalized Theta Series of a Lattice}




\author{%
  \IEEEauthorblockN{Maiara F.~Bollauf}
  \IEEEauthorblockA{University of Tartu\\
                    Narva mnt 18, 51009, Tartu, Estonia\\
                    Email: maiara.bollauf@ut.ee}
                  \thanks{This work was partially supported by the Estonian Research Council (grant PRG253).}
  \and
  \IEEEauthorblockN{Hsuan-Yin Lin}
  \IEEEauthorblockA{Simula UiB\\
                    N--5006 Bergen, Norway\\
                    Email: lin@simula.no}
}


\maketitle

\begin{abstract}
  Mimicking the idea of the generalized Hamming weight of linear codes, we introduce a new lattice invariant, the \emph{generalized theta series}. Applications range from identifying stable lattices to the lattice isomorphism problem. Moreover, we provide counterexamples for the \emph{secrecy gain conjecture} on isodual lattices, which claims that the ratio of the theta series of an isodual (and more generally, formally unimodular) lattice by the theta series of the integer lattice $\Integers^n$ is minimized at a (unique) symmetry point.
\end{abstract}


\section{Introduction}
\label{sec:introduction}

In coding theory, the generalized Hamming weight~\cite{Wei91_1} serves as a structural parameter that provides additional information beyond the minimum Hamming weight of a linear code. It has applications in the type II wiretap channel, where an eavesdropper taps $s$ out of $n$ bits of a sent message and is supposed to get the least information from it. It can also be used as a code invariant to guarantee two linear codes are not equivalent or assist in finding an equivalence if it exists. 

The theta series characterizes the (Euclidean) distance spectrum of an $n$-dimensional lattice $\Lambda$. A lattice property is said to be \emph{audible} if it can be determined by the lattice theta series, as, for example, the theta series of the dual lattice $\Lambda^\ast$ is related to the theta series of $\Lambda$ via the Jacobi's formula~\cite[p.~103]{ConwaySloane99_1}. Conway and Fung~\cite{ConwayFung09_1} asked the following question: \emph{Can you hear the shape of a lattice?} In other words, in which dimensions can there be two non-isomorphic lattices with the same theta series? It was demonstrated that one can hear the shape of $n=2$~\cite[pp.~44--45]{ConwayFung09_1} and $n=3$-dimensional lattices~\cite{Schiemann97_1}, but cannot for $n \geq 4$~\cite[pp.~42--44]{ConwayFung09_1}.

This paper contributes to the solution of this problem by providing a refined notion of audible given by a new lattice geometric invariant, the \emph{generalized theta series}. It is inspired by the generalized Hamming weight of linear codes and connects two other lattice invariants, the determinant and the theta series. In more mathematical terms, the $r$-th generalized theta series of a lattice $\Lambda$ counts the number of $r$-dimensional sublattices $\Lambda' \subseteq \Lambda$ that have the same volume. 

The first application of the generalized theta series is in finding \emph{stable lattices}, i.e., lattices such that all of its sublattices have a volume larger than or equal to one. Stable lattices have recently gained a lot of interest in connection with the \emph{reverse Minkowski theorem}~\cite{RegevStephens-Davidowitz17_1, RegevStephens-Davidowitz23_1sub}. Given the \emph{theta series ratio} $\Delta_{\Lambda}(\tau)\eqdef\nicefrac{\Theta_{\Lambda}(i \tau)}{\Theta_{\Integers^n}(i \tau)}$ of a lattice $\Lambda$, a key result in this theory is that $\Delta_{\Lambda}(\tau) \leq 1$ for all stable lattices $\Lambda$, when $\tau$ is either very small or very large~\cite{RegevStephens-Davidowitz17_1}. However, whether this inequality holds \emph{for all} $\tau>0$ remains an open problem.

In the context of wiretap channel communication, Belfiore and Solé~\cite{BelfioreSole10_1} have conjectured that the global minimum of the theta series ratio of unimodular lattices is achieved at $\tau=1$. This result is not completely demonstrated, but it is known to be true for extremal unimodular lattices~\cite{Ernvall-Hytonen12_1}, several unimodular lattices and even-dimensional Construction A unimodular lattices from binary self-dual codes in small dimensions~\cite{LinOggier12_1, LinOggier13_1}, many unimodular lattices constructed via \emph{direct-sum}~\cite{PinchakSethuraman14_1}, and Construction A and $\textnormal{A}_4$ unimodular lattices satisfying a numerical sufficient condition~\cite{BollaufLinYtrehus23_3, BollaufLinYtrehus24_1}. The conjecture was further extended to isodual~\cite{OggierSoleBelfiore16_1} and formally unimodular lattices~\cite{BollaufLinYtrehus23_3}. 

The contributions of this paper are:
\begin{itemize}
\item[i)] Consider the concepts of generalized Hamming weight and the \emph{$r$-dimensional densest sublattice problem ($r$-DSP)}~\cite{DadushMicciancio13_1}, which asks to find $r\in\{1,2,\ldots,n\}$ linearly independent vectors in an $n$-dimensional lattice $\Lambda$ that yields to the smallest volume. We propose an original lattice invariant, the \emph{generalized theta series} of a lattice $\Lambda$, which can assist in \emph{hearing the shape of a lattice}, that is, distinguishing between two non-isomorphic lattices that share the same theta series, provided their generalized theta series can be determined. Moreover, we define the \emph{$r$-th generalized Euclidean norms} of a lattice, which is a simplification of the generalized theta series. 
  
  
\item[ii)] We verify the stability of lattices via the generalized Euclidean norms of a lattice.
  
\item[iii)] We show that conjectures concerning secure communication in a Gaussian wiretap channel \emph{do not hold} for isodual lattices, as well as for formally unimodular lattices, by providing explicit counterexamples. Specifically, using techniques from the generalized theta series, we demonstrate that there exist isodual lattices such that $\Delta_{\Lambda}(\tau)>1$, and moreover, such that $\tau=1$ is not the global minimum of the theta series ratio $\Delta_{\Lambda}(\tau)$, invalidating the conjectures~\cite[Conj.~1]{OggierSoleBelfiore16_1} and~\cite[Conj.~37]{BollaufLinYtrehus23_3}, since isodual lattices are also formally unimodular.
\end{itemize}


\section{Preliminaries}
\label{sec:definitions-preliminaries}

\subsection{Notation}
\label{sec:notation}

We denote by $\Naturals$, $\Integers$, and $\Reals$ the set of naturals, integers, and reals, respectively. $[i:j]\eqdef\{i,i+1,\ldots,j\}$ for $i,j\in \Integers$, $i\leq j$. Vectors are \emph{row} vectors and boldfaced, e.g., $\vect{x}$. The all-zero vector is denoted by $\vect{0}$. Matrices and sets are represented by capital sans serif letters and calligraphic uppercase letters, respectively, e.g., $\mat{X}$ and $\set{X}$. An identity matrix $n \times n$ is denoted by $\mat{I}_n$. The inner product of two vectors is denoted by $\inner{\vect{a}}{\vect{b}}$. The natural embedding $\phi_q \colon\Integers_q^n \rightarrow \Integers^n$ is such that $\phi_q(x)$ maps each element $x\in\Integers_q$ to the corresponding integer. Let $\set{B}(s)\eqdef\{\vect{x}\in\Reals^n\colon\norm{\vect{x}}\leq s\}$ be the $n$-dimensional ball of some radius $s>0$ centered at zero.

\subsection{Lattices and Linear Codes}
\label{sec:basics_codes-lattices}

A \emph{lattice} $\Lambda\subset\Reals^n$ is a discrete additive subgroup of $\Reals^{n}$. A (full rank) lattice can also be seen as $\Lambda=\{\vect{\lambda}=\vect{u}\mat{L}_{n\times n}\colon\vect{u}\in\Integers^n\}$, where the $n$ rows of the \emph{generator matrix} $\mat{L}$ form a lattice basis in $\Reals^n$. If a lattice $\Lambda$ has generator matrix $\mat{L}$, then the lattice $\Lambda^\star\subset\Reals^n$ generated by $\trans{\bigl(\inv{\mat{L}}\bigr)}$ is called the \emph{dual lattice} of $\Lambda$. The \emph{volume} of a lattice $\Lambda$ is $\vol{\Lambda} = \ecard{\det(\mat{L})}$. A \emph{sublattice} $\Lambda'$ of a lattice $\Lambda$ is a lattice such that $\Lambda' \subseteq \Lambda$.

\begin{definition}[Theta series]
  \label{def:theta-series}
  Let $\Lambda$ be a lattice. Its theta series is given by
  \begin{IEEEeqnarray*}{c}
    \Theta_\Lambda(z) = \sum_{\vect{\lambda}\in\Lambda} q^{\norm{\vect{\lambda}}^2}
    =\sum_{\vect{\lambda}\in\Lambda}\ope^{i\pi z\norm{\vect{\lambda}}^2},
  \end{IEEEeqnarray*}
  where $q\eqdef\ope^{i\pi z}$ and $\Im{z} > 0$.
\end{definition}

Here, we will consider  $z=i\tau$ to be purely imaginary. Then, the theta series of $\Lambda$ reduces to
\begin{equation*}
  \Theta_\Lambda(i\tau)=\sum_{\vect{\lambda}\in\Lambda}\ope^{-\pi\tau\norm{\vect{\lambda}}^2}.\label{eq:theta-series_tau}
\end{equation*}

A lattice $\Lambda$ is said to be \emph{integral} if the inner product of any two lattice vectors is an integer or equivalently if and only if $\Lambda \subseteq \Lambda^\star$. An integral lattice such that $\Lambda = \Lambda^\star$ is a \emph{unimodular} lattice. A lattice that can be obtained from
its dual by a rotation or reflection is called \emph{isodual}. We say that a lattice $\Lambda$ is \emph{formally unimodular} if and only if $\Theta_{\Lambda}(z)=\Theta_{\Lambda^\star}(z)$. Notice that unimodular, isodual, and formally unimodular lattices all have volume equal to $1$.
A lattice $\Lambda$ is said to be \emph{stable} if $\vol{\Lambda}=1$ and $\vol{\Lambda'}\geq 1$ for all sublattice $\Lambda' \subseteq \Lambda$. Unimodular lattices are stable~\cite[Cor., p.~407]{Weng02_1}.

Analogous to the theta series of a lattice, a binary $[n,k]$ linear code\footnote{A binary $[n,k]$ code $\code{C}$ is a $k$-dimensional linear subspace of $\Field_2^n$. In general, codes can be defined over a Galois field $\Field_q$.} $\code{C}\subseteq\Field_2^n$ has a \emph{weight enumerator}
\begin{IEEEeqnarray*}{c}
  W_\code{C}(x,y)=\sum_{\vect{c}\in\code{C}} x^{n-\Hwt{\vect{c}}}y^{\Hwt{\vect{c}}} 
  =\sum_{w=0}^n A_w(\code{C})\, x^{n-w}y^w,\IEEEeqnarraynumspace
  \label{eq:weight-enumerator}
\end{IEEEeqnarray*}
where $A_w(\code{C})\eqdef\card{\{\vect{c}\in\code{C}\colon\Hwt{\vect{c}}=w\}}$, $w\in[0:n]$. 

We define next the \emph{generalized Hamming weight}, which characterizes the minimum weight among subcodes in binary linear codes.

\begin{definition}[Generalized Hamming weight~\cite{Wei91_1}]
  \label{def:def_GHW_Z2}
  The $r$-th generalized Hamming weight of an $[n,k]$ code $\code{C}$ is the size of the smallest support of an $r$-dimensional subcode of $\code{C}$, i.e., 
  \begin{IEEEeqnarray*}{c}
    d_{r}(\code{C})=\min\{w(\code{C}_r)\colon\code{C}_r ~ \textnormal{is an $[n,r]$ subcode of}~\code{C}
    \},
    \IEEEeqnarraynumspace
  \end{IEEEeqnarray*}
considering $w(\code{C}) = \bigl|\{i\in [1:n]\colon\exists\,\vect{c}=(c_1,\ldots,c_n) \in \code{C}~\textnormal{s.t.}~c_i \neq 0\}\bigr|$, and $r\in [1:k]$. We define $\vect{d}(\code{C})\eqdef\{d_1(\code{C}),\ldots,d_k(\code{C})\}$ as the \emph{weight hierarchy} of a code $\code{C}$ and $d_r(\code{C})$ denotes the $r$-th generalized Hamming weight of $\code{C}$.
\end{definition}

We remark that, in the literature, most results on generalized Hamming weights are established for binary $[n,k]$ codes. However, there also exist several results concerning linear codes over $\Field_q$. See, for example,~\cite{JurriusPellikaan09_1}.

The generalized Hamming weight is monotonic.

\begin{theorem}\cite[Thm.~1]{Wei91_1} 
\label{thm:order-generalized-hamming-weight}
For an $[n,k]$ linear code $\code{C}$ with $k>0$, we have that
\begin{IEEEeqnarray*}{c}
1 \leq d_1(\code{C}) < d_2(\code{C})< \cdots < d_{k}(\code{C}) \leq n.   
\end{IEEEeqnarray*}
\end{theorem}

\begin{example}
  \label{ex:generalized-hamming-weight}
  Consider two non-isometric $[6,3]$ binary codes $\code{C}_1$ and $\code{C}_2$ in~\cite[pp.~40--42]{ConwayFung09_1}, with respective generator matrices $\mat{G}^{\code{C}_1} = (\mat{I}_3 ~ \mat{B}_1)$ and $\mat{G}^{\code{C}_2} = (\mat{I}_3 ~ \mat{B}_2)$, where
  \begin{IEEEeqnarray*}{c}
    \mat{B}_1=
    \begin{pmatrix}
      1 &  0 & 0
      \\
      0 &  1 & 0
      \\
      0 &  0 & 1
    \end{pmatrix}
    \textnormal{ and }
    \mat{B}_2=
    \begin{pmatrix}
      0 &  1 & 0
      \\
      1 &  1 & 1
      \\
      0 &  1 & 0
    \end{pmatrix}.
  \end{IEEEeqnarray*}
  
  The weight hierarchies of $\code{C}_1$ and $\code{C}_2$ are 
  \begin{IEEEeqnarray*}{c}
    \vect{d}(\code{C}_1)=\{2,4,6\}\textnormal{ and }\vect{d}(\code{C}_2)=\{2,3,6\}.
  \end{IEEEeqnarray*}
  
  The distinct weight hierarchies indicate that the two codes are indeed non-isometric. However, they have the same weight enumerator 
  \begin{IEEEeqnarray*}{c}
    W_{\code{C}_1}(x,y)= W_{\code{C}_2}(x,y)=x^6+3x^4y^2+3x^2y^4+y^6,
  \end{IEEEeqnarray*}
  and thus, are said to be \emph{isospectral}.\exampleend
  
\end{example}

Lattices can be constructed from linear codes through Construction A~\cite{ConwaySloane99_1, Zamir14_1}. A $\Integers_q$ linear code $\code{C}$ of length $n$ is an additive subgroup of $\Integers_q^n$. 

\begin{definition}[{Construction A~\cite[p.~31]{Zamir14_1}}]
  \label{def:def_ConstrAq}
  Let $\code{C}$ be a $\Integers_q$ linear code, then $\ConstrAq[q]{\code{C}}\eqdef\frac{1}{\sqrt{q}}(\phi_q(\code{C}) + q\Integers^n)$ is a lattice.
\end{definition}

\subsection{Conjectures on the Theta Series}

Characterizing the theta series of a general lattice is a hard task. It has applications in many fields, being used to bound the success probability of eavesdropping a message in communication channels~\cite{BelfioreSole10_1}, or in theoretical computer science, where it is believed that the theta series of the integer lattice $\Integers^n$ maximizes the theta series of stable lattices~\cite{RegevStephens-Davidowitz24_1}.

We start by defining a particular quotient of the theta series.
\begin{definition}[{Theta series ratio~\cite{BollaufLin24_1sub}}]
  \label{def:secrecy_ratio}
  Let $\Lambda$ be a lattice with volume $\vol{\Lambda}=1$. The theta series ratio of $\Lambda$ is 
  \begin{equation*}
    \Delta_{\Lambda}(\tau)\eqdef\frac{\Theta_{\Lambda}(i \tau)}{\Theta_{\Integers^n}(i \tau)},\quad\tau\eqdef -i z>0.
    \label{eq:def_secrecy-ratio}
  \end{equation*} 
\end{definition}

Regev and Stephen-Davidowitz conjectured that $\Delta_{\Lambda}(\tau) \leq 1$ for all stable lattices~\cite{RegevStephens-Davidowitz24_1}.

\begin{conjecture}[{Upper bound on the theta series ratio for stable lattices}]
  \label{conj:RSD}
  For all stable lattices $\Lambda\subset\Reals^n$ and \emph{for all} $\tau>0$, it holds that
  \begin{equation*}
    \Theta_{\Lambda}(i \tau) \leq \Theta_{\Integers^n}(i \tau) \text{ or equivalently }\Delta_{\Lambda}(\tau)\leq 1.
  \end{equation*}
\end{conjecture}

In the context of Gaussian wiretap channel communication, the theta series ratio is also of fundamental importance since it upper bounds the error probability of an eavesdropper guessing a sent message once lattice coset encoding is performed~\cite{BelfioreSole10_1}. The original conjecture was stated for unimodular lattices.

\begin{conjecture}[{Global minimum of the theta series ratio for unimodular lattices}]
  \label{conj:conjecture_BelfioreSole_unimodular-lattices}
  The theta series ratio of a unimodular lattice $\Lambda$ achieves its global minimum at $\tau=1$, i.e.,
  \begin{equation*}
    \argmin_{\tau>0}\Delta_{\Lambda}(\tau)=1.
    \label{eq:global-minimum_unimodular-lattices}
  \end{equation*}
\end{conjecture}
Later on, the same conjecture was extended to isodual~\cite[Conj.~1]{OggierSoleBelfiore16_1} and formally unimodular lattices~\cite[Conj.~37]{BollaufLinYtrehus23_3}. Given that formally unimodular lattices are also isodual, we will focus on isodual lattices from this point onward. Nevertheless, the same conclusions apply to formally unimodular lattices.  

\begin{figure}[t!]
    \centering
    \input{\Figs/good_secrecy_gain_bordered_double_circulant_n6.tex} 
    \caption{Theta series ratios as a function of $\tau>0$ for several \emph{isodual} lattices that both satisfy Conjectures~\ref{conj:RSD} and~\ref{conj:conjecture_BelfioreSole_unimodular-lattices}. Observe that $\Delta_{\Lambda}(\tau)\leq 1$ for all $\tau>0$ and $\argmin_{\tau>0}\Delta_{\Lambda}(\tau)=1$.}
    \label{fig:good_secrecy_gain_bordered_double_circulant_n6}
  \end{figure}
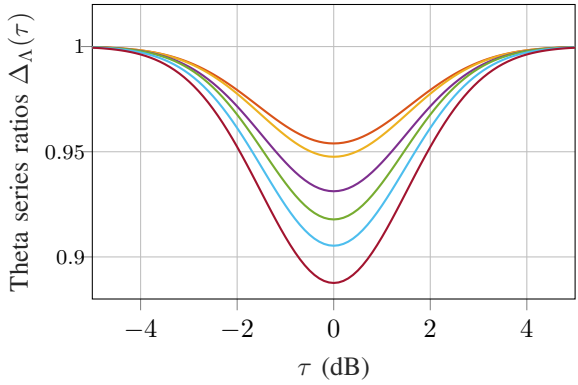

Fig.~\ref{fig:good_secrecy_gain_bordered_double_circulant_n6} illustrates several \emph{typical} isodual lattices that simultaneously satisfy Conjectures~\ref{conj:RSD} and~\ref{conj:conjecture_BelfioreSole_unimodular-lattices}.

The argument for the minimization of the theta series ratio, relies on the concept of \emph{weak secrecy gain}, which is simply the theta series ratio evaluated at a symmetry point $\tau_0$, i.e., $\Delta_{\Lambda}(\tau_0)$, where $\tau_0$ is such that for all $\tau>0$,
\begin{equation*}
  \Delta_{\Lambda}(\tau_0 \cdot \tau) = \Delta_{\Lambda}(\nicefrac{\tau_0}{\tau}).
\end{equation*}

In~\cite[Conj.~1]{OggierSoleBelfiore16_1}, the claim is that, given an isodual lattice $\Lambda$, the global minimum of its theta series ratio is achieved at the symmetry point $\tau_0=1$. We will refer to this formulation as the \emph{secrecy gain conjecture} for isodual lattices.

\section{Generalized Theta Series}
\label{sec:generalized_theta_series}

Inspired by the concept of generalized Hamming weight where it determines the smallest ``weight'' of an $r$-dimensional subcode of $\code{C}$, we review an analogous notion in the lattice context: a generalization of the \emph{shortest vector problem} (SVP) in a lattice $\Lambda$, namely the $r$-DSP~\cite{DadushMicciancio13_1}.
\begin{definition}[$r$-DSP]
  \label{def:r-DSP}
  Consider a lattice $\Lambda\subseteq\Reals^n$. Find $r$ linearly independent lattice vectors $\{\vect{a}_1,\vect{a}_2,\ldots, \vect{a}_r\}\subseteq\Lambda$ such that it generates a sublattice achieving the smallest possible volume $\det(\mat{A}\trans{\mat{A}})$, where $\trans{\mat{A}}= [\trans{\vect{a}}_1,\ldots,\trans{\vect{a}}_r]$, $r \in [1:n]$.
\end{definition}

Let $\lambda_1$ be the length of the shortest nonzero vector of a lattice $\Lambda$. The main theoretical finding in~\cite{DadushMicciancio13_1} is the realization that the $r$-DSP solution either contains the lattice shortest vectors or one can efficiently generate a list of $\ebigOf{r}^n$ lattice vectors with length at most $r\lambda_1$ that contains the $r$-DSP solution as a subset.
\begin{lemma}[{\cite[Lemma~3.1]{DadushMicciancio13_1}}]
  \label{lem:determining_rDSP}
  Consider an $n$-dimensional lattice $\Lambda$. A minimum-volume sublattice either contains all lattice vectors of length $\lambda_1$, or it contains a set of $r$ linearly independent vectors, each of length at most $r\lambda_1$.
\end{lemma}
This result serves as the main motivation to define an analogous notion, the ``generalized weight enumerator'' for lattices.

\begin{definition}[Generalized Theta Series]
  \label{def:generalized-theta-series}
  Consider a lattice $\Lambda\subset\Reals^{n}$. Its $r$-th generalized theta series is
  \begin{equation}
    \Theta^{(r)}_{\Lambda}(z) = \sum_{\substack{\{\vect{a}_i\}_{i=1}^r\subseteq\Lambda\cap \set{B}(r\lambda_1^{(m)})\colon\\[0.75mm]\erank{\mat{A}}=r,}}q^{\det(\mat{A}\trans{\mat{A}})}, \label{eq:conj_generalized-theta-series}
  \end{equation}
  where $\trans{\mat{A}}= [\trans{\vect{a}}_1,\ldots,\trans{\vect{a}}_r]$, $r \in [1:n]$, $q\eqdef\ope^{i \pi z}$ and $\Im{z} > 0$. Here, $\lambda_1^{(m)}$ refers to the length of the $m$-th shortest vector in $\Lambda$, which is equal to $\sqrt{\mu_m}$ of the $m$-th term of the first generalized theta series $\Theta^{(1)}_{\Lambda}(z)=\sum_{m} N_m q^{\mu_m}$. Moreover, $\lambda_1^{(1)}=\lambda_1$ is equivalent to the first \emph{successive minima} of a lattice~\cite{MicciancioGoldwasser02_1}.
\end{definition}

\begin{figure}[t!]
  \centering
  \input{\Figs/gts-region.tex} 
  \caption{Geometric interpretation of the generalized theta series.}
  \label{fig:ex_fundamental_3_7}
\end{figure}
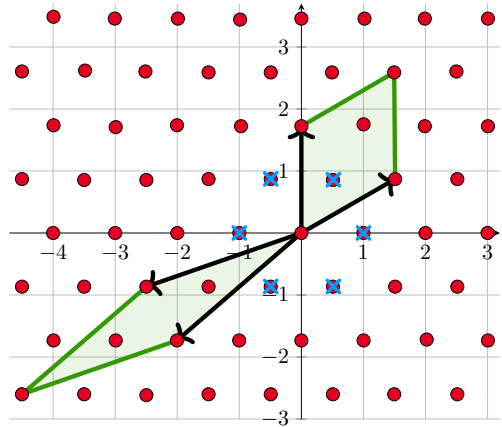

Observe that the definition of generalized theta series does not take into account the ordering. In other words, the lattice generated by any permutation of the vectors $\{\vect{a}_1,\ldots,\vect{a}_r\}$ is considered just once in the exponent of~\eqref{eq:conj_generalized-theta-series}. 


\begin{remark}~
  \label{rem:remark1}
  \begin{enumerate}
  \item $\Theta_{\Lambda}(z)=1+\Theta^{(1)}_{\Lambda}(z)$.
  \item The set $\{\vect{a}_1,\ldots,\vect{a}_r\}\subseteq\Lambda$ consisting of $r$ linearly independent lattices vectors generates an $r$-dimensional sublattice $\Lambda' \subseteq\Lambda\subset\Reals^n$. Its volume is $\vol{\Lambda'}=\sqrt{\det(\mat{A}\trans{\mat{A}})}$ where $\trans{\mat{A}}= [\trans{\vect{a}}_1,\ldots,\trans{\vect{a}}_r]$, $r \in [1:n]$.
  \end{enumerate}
\end{remark}

\begin{example}
  Consider the hexagonal lattice $\textnormal{A}_2$, with basis $\{(1,0),(\nicefrac{1}{2},\nicefrac{\sqrt{3}}{2})\}$. From Definition~\ref{def:generalized-theta-series}, we get that
  \begin{IEEEeqnarray*}{rCl}
    \Theta^{(1)}_{\textnormal{A}_2}(z)& = &6q + 6q^3 + 6q^4 + 12q^7 + 6q^9 + \cdots,
    \\
    \Theta^{(2)}_{\textnormal{A}_2}(z)& = & 36 q^{\nicefrac{3}{4}} + 156 q^{3}+168 q^{\nicefrac{27}{4}} + 380 q^{12} + \cdots.
  \end{IEEEeqnarray*}
  
  Geometrically, given the $m$-th term $N_m q^{\mu_m}$ of the generalized theta series $\Theta^{(r)}_{\Lambda}(z)$, the exponent $\mu_m$ corresponds to the $m$-th smallest volume of the fundamental region of a sublattice generated by $r$ linearly independent lattice vectors within a ball of radius $r\lambda_1^{(m)}$. The integer $N_m$ indicates how many sets of $r$ linearly independent vectors in $\Lambda\cap\set{B}\bigl(r\lambda_1^{(m)}\bigr)$ have such volume. The blue crosses in Fig.~\ref{fig:ex_fundamental_3_7} illustrate the six vectors of length one in the first term of $\Theta^{(1)}_{\textnormal{A}_2}(z)$, while the green regions (with the same area) are generated by two sets of vectors that contribute to the term $168 q^{\nicefrac{27}{4}}$ in $\Theta^{(2)}_{\textnormal{A}_2}(z)$.\exampleend
\end{example}

Apart from enumerating the $r$-dimensional volumes for $\Lambda$, we simply 
define the corresponding \emph{generalized Euclidean norm} for a lattice $\Lambda$.
\begin{definition}[{$r$-th Generalized Euclidean Norm/$r$-Dimensional Minimum Sublattice Volume}]
  \label{def:generalized-euclidean-norm}
  The $r$-th generalized Euclidean norms are the minimum exponents defined in~\eqref{eq:conj_generalized-theta-series} for all $r\in [1:n]$ and, 
  \begin{equation*}
    \nu_{r}(\Lambda) = \min\{\det(\mat{A}\trans{\mat{A}}): \{\vect{a}_i\}_{i=1}^r\subseteq\Lambda\textnormal{ and }\erank{\mat{A}}=r\}.
  \end{equation*}
  Moreover, the norm hierarchy is defined as $\vect{\nu}(\Lambda)=\{\nu_{r}(\Lambda)\colon r\in [1:n]\}$.
\end{definition}
Note that this also corresponds to the \emph{determinantal minima}~\cite[Def.~3.13]{Dadush18_1sub} in the computer science literature.

\begin{remark}
  \label{rem:SSV_r_1-n}
  The generalized Euclidean norm simply captures the leading exponent in the generalized theta series of a lattice, and the first term of the generalized theta series $\Theta^{(r)}_{\Lambda}(z)$ resolves the $r$-DSP problem. It follows from Definition~\ref{def:generalized-euclidean-norm} that we have $\nu_{1}(\Lambda) = \lambda^2_1$ of the lattice $\Lambda$ and $\nu_{n}(\Lambda) = \vol{\Lambda}^2$.
\end{remark}

\section{Properties of the\\ $r$-th Generalized Euclidean Norm}
\label{sec:properties_rth-GEN}

We now present a property related to the $r$-th generalized Euclidean norm for equivalent lattices.
\begin{proposition}
  \label{prop:gen-equivalent-lattices} 
  Consider two equivalent lattices $\Lambda, \widebar{\Lambda} \subseteq \Reals^n$, i.e., $\mat{L}_{\widebar{\Lambda}} = \alpha\mat{L}_{\Lambda}\mat{Q}$ for some $\alpha\neq 0$ and an orthogonal matrix $\mat{Q} \in \Reals^{n \times n}$. Then, $ \nu_r(\widebar{\Lambda}) = \alpha^{2r}\nu_r(\Lambda)$ for all $r\in [1:n]$.
\end{proposition}

\begin{IEEEproof}
  Consider $\{\widebar{\vect{a}}_i\}_{i=1}^r \subseteq \widebar{\Lambda}$, $\trans{\widebar{\mat{A}}}= [\trans{\widebar{\vect{a}}_1},\ldots,\trans{\widebar{\vect{a}}_r}]$, $r\in [1:n]$, and $\erank{\widebar{\mat{A}}}=r$. Observe that for a fixed $i$ and $\widebar{\vect{a}}_i \in \widebar{\Lambda}$, we have $\widebar{\vect{a}}_i=\vect{u}_i\mat{L}_{\widebar{\Lambda}}=\alpha \vect{u}_i \mat{L}_{\Lambda}\mat{Q}$, where $\vect{u}_i \in \Integers^n$. Therefore, the \emph{Gram matrix} $\widebar{\mat{A}}\trans{\widebar{\mat{A}}}$~\cite[p.~101]{ConwaySloane99_1} will have elements of the form
  \begin{IEEEeqnarray*}{rCl}
    \langle\widebar{\vect{a}}_i, \widebar{\vect{a}}_j \rangle & = & \langle \alpha \vect{u}_i \mat{L}_{\Lambda}\mat{Q}, \alpha \vect{u}_j \mat{L}_{\Lambda}\mat{Q} \rangle = \alpha^2 \langle \vect{u}_i \mat{L}_{\Lambda}, \vect{u}_j \mat{L}_{\Lambda} \rangle \\
    & = & \alpha^2 \langle \vect{a}_i, \vect{a}_j \rangle,
  \end{IEEEeqnarray*}
  for $i,j \in [1:r]$, $\vect{a}_i, \vect{a}_j \in \Lambda$. Since $\widebar{\mat{A}}\trans{\widebar{\mat{A}}}$ and $\mat{A}\trans{\mat{A}}$ are $r \times r$ matrices for a fixed rank $r$, we can conclude that
  \begin{equation*}
    \det(\widebar{\mat{A}}\trans{\widebar{\mat{A}}}) = \alpha^{2r}\det(\mat{A}\trans{\mat{A}}).
  \end{equation*}
  This completes the proof.
\end{IEEEproof}

\begin{example}
  \label{ex:example_D4-equivalent}
  Consider the $\textnormal{D}_4$ lattice~\cite[p.~9]{ConwaySloane99_1} generated by the following generator matrix
  \begin{equation*}
    \mat{L}_{\textnormal{D}_4} =
    \Scale[0.80]{\begin{pmatrix}
      2 & 0 & 0 & 0 \\
      1 & 1 & 0 & 0 \\
      1 & 0 & 1 & 0 \\
      1 & 0 & 0 & 1
    \end{pmatrix}},  
  \end{equation*}
  and another lattice $\widebar{\textnormal{D}}_4$ with generator matrix $\mat{L}_{\widebar{\textnormal{D}}_4}=\frac{1}{\sqrt{2}}\mat{L}_{\textnormal{D}_4}\mat{Q}$, which is equivalent to $\textnormal{D}_4$ and
  \begin{equation*}
    \mat{Q}=\frac{1}{\sqrt{2}}
    \Scale[0.80]{\begin{pmatrix}
      1 & 1 & 0 & 0 \\
      1 & -1 & 0 & 0 \\
      0 & 0 & 1 & 1 \\
      0 & 0 & 1 & -1
    \end{pmatrix}}
  \end{equation*}
  is an orthogonal matrix. As a result, we numerically compute the norm hierarchy $\vect{\nu}(\textnormal{D}_4) = (2,3,4,4)$ based on Definition~\ref{def:generalized-euclidean-norm}, and Proposition~\ref{prop:gen-equivalent-lattices} indicates that $\vect{\nu}(\widebar{\textnormal{D}}_4) = (1,\nicefrac{3}{4},\nicefrac{1}{2},\nicefrac{1}{4})$.\exampleend
\end{example}

\section{Applications}
\label{sec:applications}

\subsection{Stability of Lattices}
\label{sec:stability_lattice}

The generalized Euclidean norms naturally can identify whether a lattice is stable, which we address next.

To the best of our knowledge, the most efficient algorithm to compute the $r$-DSP has running time at most $r^{\ebigOf{r n}}$, which is presented in~\cite{DadushMicciancio13_1}.
To verify whether a lattice is stable, one must ensure that for any $\Lambda' \subseteq\Lambda$, $\vol{\Lambda'}\geq 1$. Thus, Lemma~\ref{lem:determining_rDSP} can be used to verify the stability of a lattice computationally. In the following examples, we provide three concrete evidence demonstrating the fact that Construction A lattices obtained from codes over $\Integers_q$ are not necessarily stable. We begin with an example based on the binary Construction A lattice, building upon Example~\ref{ex:generalized-hamming-weight}.

\begin{example}
  \label{ex:generalized-theta-series_n6k3}
  Consider the corresponding Construction A lattices $\ConstrAq[2]{\code{C}_1}$ and $\ConstrAq[2]{\code{C}_2}$, obtained from $\code{C}_1$ and $\code{C}_2$ as in Example~\ref{ex:generalized-hamming-weight}, respectively. Using~\cite[Algorithm 1]{DadushMicciancio13_1} we get
  \begin{IEEEeqnarray*}{c}
    \vect{\nu}(\Lambda_1)=\{1,1,1,1,1,1\},~\vect{\nu}(\Lambda_2)=\{1,3/4,1/2,3/4,1,1\},
  \end{IEEEeqnarray*}
  which shows that $\ConstrAq[2]{\code{C}_2}$ is not stable as there exists an $2$-dimensional $\Lambda' \subseteq\ConstrAq[2]{\code{C}_2}$ with $\vol{\Lambda'}<1$.

  In fact, using Definition~\ref{def:generalized-theta-series}, with an extensive computation, we get
  \begin{IEEEeqnarray*}{rCl}
    \Theta^{(1)}_{\ConstrAq[2]{\code{C}_1}}(z)& = &\Theta^{(1)}_{\ConstrAq[2]{\code{C}_2}}(z)\nonumber
    \\
    & = & 12 q^{1}+ 60 q^{2} + 160 q^{3}+ 252 q^{4}+\cdots,
    \\[1mm]
    \Theta^{(2)}_{\ConstrAq[2]{\code{C}_1}}(z)& = & 300 q^{1}+ \bm{3936} q^{2}+ \bm{9984} q^{3}+\cdots,
    \\
    \Theta^{(2)}_{\ConstrAq[2]{\code{C}_2}}(z)& = &144 q^{\nicefrac{3}{4}}+ \bm{92} q^{1} + \bm{1920} q^{\nicefrac{7}{4}}+\cdots.
  \end{IEEEeqnarray*}
  It is worth mentioning that the boldfaced coefficients cannot be guaranteed by Lemma~\ref{lem:determining_rDSP}, as they do not correspond to the minimum sublattice volume. Here, we simply obtained the values numerically. Efficient and accurate computation of the exact volume of non-minimum sublattice is an interesting problem, which we leave for future investigation.

  Furthermore, our findings indicate that the two Construction A lattices $\ConstrAq[2]{\code{C}_1}$ and $\ConstrAq[2]{\code{C}_2}$ are non-isometric (even though they have the same first generalized theta series), which partially addresses the question raised in~\cite{ConwayFung09_1}: ``Can one can hear the shape of a lattice?'' Thus, it appears that we can indeed ``hear'' the shapes of lattices through this newly introduced definition of the \emph{generalized theta series} for lattices.\exampleend
\end{example}

Similarly, as the generalized Hamming weight can be used to distinguish equivalent codes, the generalized theta series serves as a geometric invariant that can help determine whether two lattices are isometric, which is the hard problem behind the \emph{Lattice Isomorsphim Problem} (LIP)~\cite{DucasvanWoerden22_1}. We emphasize, however, that this does not necessarily pose a threat to cryptographic schemes based on the LIP, since computing the $r$-th generalized Euclidean norm or generalized theta series remains computationally expensive and, therefore, impractical for general lattices.

\subsection{Conjectures Do Not Hold for Isodual Lattices!}

The following counterexample disproves the secrecy gain conjecture for isodual lattices~\cite[Conj.~1]{OggierSoleBelfiore16_1}.


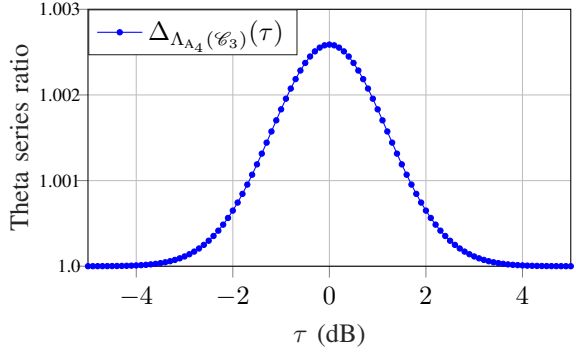
\begin{figure}[t!]
  \centering
  \input{\Figs/bad_secrecy_gain_bordered_double_circulant_n6.tex} 
  \caption{Theta series ratio as a function of $\tau>0$ in Example~\ref{ex:counter-example_Conjectures1-2}. Observe that $\Delta_{\ConstrAq[4]{\code{C}_3}}(\tau)>1$ for all $\tau>0$.}
  \label{fig:bad_secrecy_gain_bordered_double_circulant_n6}
\end{figure}

\begin{example}
  \label{ex:counter-example_Conjectures1-2}
  Consider a $\Integers_4$-linear code generated by
  \begin{equation*}
    \mat{G}^{\code{C}_3} =
    \Scale[0.80]{\begin{pNiceMatrix}
      1 & 0 & 0 & 1 & 2 & 2
      \\
      0 & 1 & 0 & 2 & 0 & 2
      \\
      0 & 0 & 1 & 2 & 2 & 0
    \end{pNiceMatrix}}.
  \end{equation*}
  Applying~\cite[Algorithm 1]{DadushMicciancio13_1} we obtain
  \begin{IEEEeqnarray*}{c}
    \vect{\nu}\bigl(\ConstrAq[4]{\code{C}_3}\bigr)=\{1,\nicefrac{3}{4},\nicefrac{1}{2},\nicefrac{3}{4},1,1\},
  \end{IEEEeqnarray*}
  and thus the lattice $\ConstrAq[4]{\code{C}_3}$ is \emph{not stable}. Moreover, its generalized theta series is given by
  \begin{IEEEeqnarray*}{rCl}
    \Theta^{(1)}_{\ConstrAq[4]{\code{C}_3}}(z)& = &12 q^{1} + 16 q^{\nicefrac{7}{4}} + 8 q^2 + 32 q^{\nicefrac{9}{4}} + \cdots,
    \\
    \Theta^{(2)}_{\ConstrAq[4]{\code{C}_3}}(z)& = &144q^{\nicefrac{3}{4}} + \bm{124} q^{1}+\bm{144} q^{\nicefrac{3}{2}}+\cdots.
  \end{IEEEeqnarray*}  
  Note that $\ConstrAq[4]{\code{C}_3}=\frac{1}{2}(\phi_4(\code{C}_3)+4\mathbb{Z}^n)$ and $\code{C}$ is an isodual bordered double circulant code~\cite[Lemma 2.4]{BachocGulliverHarada00_1}, thus $\ConstrAq[4]{\code{C}_3}$ is isodual, \cite[p.~378]{HuffmanPless03_1},~\cite[Sec.~III-B]{BollaufLinYtrehus24_1}. Its theta series ratio is illustrated in Fig.~\ref{fig:bad_secrecy_gain_bordered_double_circulant_n6}, and $\Delta_{\ConstrAq[4]{\code{C}_3}}(1) \approx 1.0026>1$ (more details about the calculation can be found in~\cite{BollaufLinYtrehus24_1}), which demonstrates that Conjecture~\ref{conj:RSD} is not true for isodual lattices. However, this does not disprove the conjecture since isodual lattices are not necessarily stable, as shown in this example. We also observe that, although its theta series ratio exhibits one symmetry point, it attains its \emph{maximum} at $\tau=1$, rather than the minimum. Therefore, Conjecture~\ref{conj:conjecture_BelfioreSole_unimodular-lattices} does not hold as well. This invalidates the current formulation of the secrecy gain conjecture for isodual lattices~\cite[Conj.~1]{OggierSoleBelfiore16_1}, and consequently, its generalization to formally unimodular lattices presented in~\cite[Conj.~37]{BollaufLinYtrehus23_3}. \exampleend

  \end{example}

  Next, we provide another compelling example that violates Conjecture~\ref{conj:RSD} while satisfying Conjecture~\ref{conj:conjecture_BelfioreSole_unimodular-lattices}.

\begin{figure}[t!]
    \centering
    \input{\Figs/conj1_secrecy_gain_bordered_double_circulant_n6.tex} 
    \caption{Theta series ratio as a function of $\tau>0$ for a $\ConstrAq[4]{\code{C}_4}$ lattices that satisfy Conjecture~\ref{conj:conjecture_BelfioreSole_unimodular-lattices}. However, it can be observed that it does not satisfy Conjecture \ref{conj:RSD}; that is, there exists some $\tau>0$ such that $\Delta_{\ConstrAq[4]{\code{C}_4}}(\tau)>1$.}
    \label{fig:conj1_secrecy_gain_bordered_double_circulant_n6}
  \end{figure}
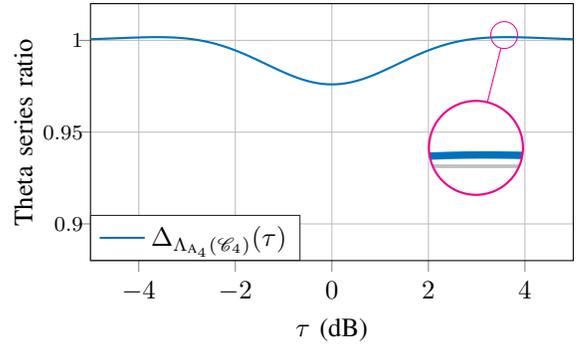

  \begin{example}
    \label{ex:counter-example_Conjectures1}
    Consider a $\Integers_4$-linear code generated by
    \begin{equation*}
      \mat{G}^{\code{C}_4} =
      \Scale[0.80]{\begin{pNiceMatrix}
        1 & 0 & 0 & 0 & 1 & 1
        \\
        0 & 1 & 0 & 1 & 0 & 2
        \\
        0 & 0 & 1 & 1 & 2 & 0
      \end{pNiceMatrix}}.
    \end{equation*}

    Applying~\cite[Algorithm 1]{DadushMicciancio13_1} we get
    \begin{equation*}
      \vect{\nu}\bigl(\ConstrAq[4]{\code{C}_4}\bigr)=\{0.75,0.88,0.77,0.88,0.75,1\},
    \end{equation*}
    which indicates that $\ConstrAq[4]{\code{C}_4}$ is not stable.
    
    We demonstrate the theta series ratio in Fig.~\ref{fig:conj1_secrecy_gain_bordered_double_circulant_n6}. As shown, the theta series ratio clearly reaches its minimum at $\tau=1$, thereby satisfying Conjecture~\ref{conj:conjecture_BelfioreSole_unimodular-lattices}. Nevertheless, there exist regions of $\tau$ where the theta series ratio remains strictly greater than $1$, revealing that Conjecture~\ref{conj:RSD} does not hold for this isodual lattice.
  \end{example}

\section{Conclusion}
\label{sec:conclusion}

We have presented a new lattice invariant, the generalized theta series. It characterizes the volume of lattices generated by $r$ linearly independent lattice vectors, with $r \in [1:n]$. In terms of applications, calculating the generalized theta series of a lattice solves the $r$-DSP, serves as an auxiliary tool to decide whether two lattices are isomorphic, and can be used to find stable lattices. In this work, we have applied this new lattice property to find counterexamples for a decade-long conjecture about the secrecy gain of isodual (and more recently, formally self-dual) lattices. Moving forward, we want to demonstrate further properties of the generalized theta series, find relations through the Jacobi theta functions~\cite[pp.~102--105]{ConwaySloane99_1} to speed up its rather costly calculations. We also aim to further investigate the relationship between the $r$-th generalized Hamming weights of a code and the $r$-th generalized Euclidean norms of the corresponding lattices derived from the code.

\section*{Acknowledgment}
The authors would like to thank Russell W.\ F.\ Lai and Wessel van Woerden for their comments on defining the generalized theta series of a lattice.






\IEEEtriggeratref{13}

\bibliographystyle{IEEEtran}
\bibliography{defshort1,biblioHY}









\end{document}

%% file: figs/good_secrecy_gain_bordered_double_circulant_n6.tex
%
%
\definecolor{mycolor1}{rgb}{0.00000,0.44700,0.74100}%
\definecolor{mycolor2}{rgb}{0.85000,0.32500,0.09800}%
\definecolor{mycolor3}{rgb}{0.92900,0.69400,0.12500}%
\definecolor{mycolor4}{rgb}{0.49400,0.18400,0.55600}%
\definecolor{mycolor5}{rgb}{0.46600,0.67400,0.18800}%
\definecolor{mycolor6}{rgb}{0.30100,0.74500,0.93300}%
\definecolor{mycolor7}{rgb}{0.63500,0.07800,0.18400}%
\begin{tikzpicture}

\begin{axis}[%
width=8cm,
height=5.50cm,
tick align=outside,
tick pos=left,
xmin=-5,
xmax=5,
xlabel style={font=\color{white!15!black}},
xlabel={$\tau$ (dB)},
ymin=0.88,
ymax=1.02,
y tick label style={font=\footnotesize, xshift=5pt},  
ylabel style={font=\color{white!15!black}, yshift=-0.10cm},
ylabel={Theta series ratios $\Delta_{\Lambda}(\tau)$},
xmajorgrids,
ymajorgrids,
legend style={legend cell align=left, align=left, draw=white!15!black}
]

\addplot [color=mycolor2, thick]
  table[row sep=crcr]{%
-5	0.999744291454855\\
-4.9	0.999682185093017\\
-4.8	0.999607201355365\\
-4.7	0.999517191205839\\
-4.6	0.999409756663128\\
-4.5	0.999282242932373\\
-4.4	0.999131734508446\\
-4.3	0.998955056083093\\
-4.2	0.998748779104343\\
-4.1	0.998509234830061\\
-4	0.998232534684827\\
-3.9	0.997914598666861\\
-3.8	0.997551192456641\\
-3.7	0.997137973748496\\
-3.6	0.996670548159232\\
-3.5	0.996144534862814\\
-3.4	0.995555641857895\\
-3.3	0.994899750497213\\
-3.2	0.994173008598287\\
-3.1	0.993371931118492\\
-3	0.992493507022116\\
-2.9	0.991535310601529\\
-2.8	0.990495615150776\\
-2.7	0.989373506541024\\
-2.6	0.98816899392848\\
-2.5	0.986883114552864\\
-2.4	0.985518029375351\\
-2.3	0.984077106175856\\
-2.2	0.982564986696625\\
-2.1	0.98098763449624\\
-2	0.979352360376373\\
-1.9	0.977667822569891\\
-1.8	0.975943999335212\\
-1.7	0.974192132184508\\
-1.6	0.972424638672442\\
-1.5	0.970654994471284\\
-1.4	0.968897585334943\\
-1.3	0.967167530480431\\
-1.2	0.965480479857822\\
-1.1	0.963852388702913\\
-1	0.962299273633077\\
-0.899999999999999	0.960836955318918\\
-0.8	0.959480793407697\\
-0.7	0.958245419858048\\
-0.6	0.957144477144499\\
-0.5	0.956190367886839\\
-0.399999999999999	0.955394022343636\\
-0.3	0.954764689880194\\
-0.199999999999999	0.954309759986207\\
-0.0999999999999996	0.954034617693408\\
0	0.953942537351658\\
0.0999999999999996	0.954034617693408\\
0.199999999999999	0.954309759986207\\
0.3	0.954764689880193\\
0.399999999999999	0.955394022343636\\
0.5	0.956190367886839\\
0.6	0.957144477144499\\
0.7	0.958245419858048\\
0.8	0.959480793407697\\
0.899999999999999	0.960836955318918\\
1	0.962299273633077\\
1.1	0.963852388702914\\
1.2	0.965480479857821\\
1.3	0.967167530480431\\
1.4	0.968897585334943\\
1.5	0.970654994471284\\
1.6	0.972424638672441\\
1.7	0.974192132184509\\
1.8	0.975943999335211\\
1.9	0.97766782256989\\
2	0.979352360376373\\
2.1	0.980987634496243\\
2.2	0.982564986696627\\
2.3	0.984077106175858\\
2.4	0.985518029375348\\
2.5	0.986883114552867\\
2.6	0.988168993928482\\
2.7	0.989373506541024\\
2.8	0.990495615150776\\
2.9	0.99153531060153\\
3	0.992493507022117\\
3.1	0.993371931118493\\
3.2	0.994173008598284\\
3.3	0.994899750497213\\
3.4	0.995555641857893\\
3.5	0.99614453486281\\
3.6	0.996670548159229\\
3.7	0.997137973748497\\
3.8	0.997551192456638\\
3.9	0.997914598666857\\
4	0.998232534684825\\
4.1	0.998509234830066\\
4.2	0.998748779104343\\
4.3	0.998955056083089\\
4.4	0.999131734508449\\
4.5	0.999282242932375\\
4.6	0.999409756663124\\
4.7	0.999517191205844\\
4.8	0.999607201355361\\
4.9	0.999682185093016\\
5	0.999744291454862\\
};

\addplot [color=mycolor3, thick]
  table[row sep=crcr]{%
-5	0.99974204391346\\
-4.9	0.999679067189055\\
-4.8	0.99960291177004\\
-4.7	0.999511337481936\\
-4.6	0.999401832051888\\
-4.5	0.999271598693771\\
-4.4	0.999117547140059\\
-4.3	0.99893628899082\\
-4.2	0.998724138300181\\
-4.1	0.998477118354466\\
-4	0.998190975610188\\
-3.9	0.997861201750007\\
-3.8	0.997483064776396\\
-3.7	0.997051649991757\\
-3.6	0.996561911605879\\
-3.5	0.996008735562894\\
-3.4	0.995387013986811\\
-3.3	0.994691731404388\\
-3.2	0.993918062615001\\
-3.1	0.993061481738773\\
-3	0.992117881588388\\
-2.9	0.99108370208034\\
-2.8	0.989956065934742\\
-2.7	0.988732919418749\\
-2.6	0.987413175380619\\
-2.5	0.985996855316164\\
-2.4	0.984485226727649\\
-2.3	0.982880931601085\\
-2.2	0.981188101468493\\
-2.1	0.979412454266441\\
-2	0.977561368080895\\
-1.9	0.975643926910879\\
-1.8	0.973670933816142\\
-1.7	0.971654887259007\\
-1.6	0.969609917122239\\
-1.5	0.967551677788075\\
-1.4	0.965497196791386\\
-1.3	0.96346467889167\\
-1.2	0.961473266908759\\
-1.1	0.959542762285682\\
-1	0.95769331001495\\
-0.899999999999999	0.955945054215879\\
-0.8	0.95431777219679\\
-0.7	0.95283049618926\\
-0.6	0.951501133016911\\
-0.5	0.950346092680423\\
-0.399999999999999	0.949379937140565\\
-0.3	0.94861506041773\\
-0.199999999999999	0.948061410480416\\
-0.0999999999999996	0.947726262274074\\
0	0.947614049681878\\
0.0999999999999996	0.947726262274073\\
0.199999999999999	0.948061410480416\\
0.3	0.94861506041773\\
0.399999999999999	0.949379937140565\\
0.5	0.950346092680423\\
0.6	0.951501133016911\\
0.7	0.952830496189261\\
0.8	0.954317772196789\\
0.899999999999999	0.955945054215879\\
1	0.957693310014949\\
1.1	0.959542762285682\\
1.2	0.961473266908759\\
1.3	0.96346467889167\\
1.4	0.965497196791386\\
1.5	0.967551677788075\\
1.6	0.969609917122238\\
1.7	0.971654887259008\\
1.8	0.973670933816141\\
1.9	0.975643926910877\\
2	0.977561368080895\\
2.1	0.979412454266444\\
2.2	0.981188101468495\\
2.3	0.982880931601087\\
2.4	0.984485226727646\\
2.5	0.985996855316167\\
2.6	0.987413175380621\\
2.7	0.98873291941875\\
2.8	0.989956065934742\\
2.9	0.99108370208034\\
3	0.992117881588389\\
3.1	0.993061481738775\\
3.2	0.993918062614998\\
3.3	0.994691731404388\\
3.4	0.995387013986808\\
3.5	0.99600873556289\\
3.6	0.996561911605876\\
3.7	0.997051649991759\\
3.8	0.997483064776393\\
3.9	0.997861201750003\\
4	0.998190975610187\\
4.1	0.998477118354472\\
4.2	0.998724138300181\\
4.3	0.998936288990815\\
4.4	0.999117547140063\\
4.5	0.999271598693774\\
4.6	0.999401832051883\\
4.7	0.999511337481941\\
4.8	0.999602911770036\\
4.9	0.999679067189053\\
5	0.999742043913467\\
};

\addplot [color=mycolor4, thick]
  table[row sep=crcr]{%
-5	0.999716044976092\\
-4.9	0.99964503862238\\
-4.8	0.999558684759025\\
-4.7	0.999454249521929\\
-4.6	0.999328639013577\\
-4.5	0.999178377034736\\
-4.4	0.998999586448058\\
-4.3	0.998787975388991\\
-4.2	0.998538829650163\\
-4.1	0.998247012658317\\
-4	0.997906974531463\\
-3.9	0.99751277174089\\
-3.8	0.997058098900339\\
-3.7	0.996536334154977\\
-3.6	0.995940599537487\\
-3.5	0.995263837489751\\
-3.4	0.994498904508612\\
-3.3	0.993638682556733\\
-3.2	0.992676208479694\\
-3.1	0.991604821185583\\
-3	0.99041832577394\\
-2.9	0.989111173150996\\
-2.8	0.987678652946617\\
-2.7	0.98611709676912\\
-2.6	0.984424088017146\\
-2.5	0.982598673639327\\
-2.4	0.98064157242517\\
-2.3	0.97855537366321\\
-2.2	0.976344719359026\\
-2.1	0.97401646271428\\
-2	0.971579795278059\\
-1.9	0.969046335143018\\
-1.8	0.966430168815762\\
-1.7	0.963747839981461\\
-1.6	0.961018279332585\\
-1.5	0.9582626709513\\
-1.4	0.955504252416134\\
-1.3	0.952768047815288\\
-1.2	0.950080535137753\\
-1.1	0.94746925200126\\
-1	0.94496234626337\\
-0.899999999999999	0.942588080630007\\
-0.8	0.940374302793026\\
-0.7	0.938347894757354\\
-0.6	0.936534216724606\\
-0.5	0.934956562061311\\
-0.399999999999999	0.933635640396235\\
-0.3	0.932589105693075\\
-0.199999999999999	0.931831145200572\\
-0.0999999999999996	0.931372143503303\\
0	0.931218433538229\\
0.0999999999999996	0.931372143503303\\
0.199999999999999	0.931831145200572\\
0.3	0.932589105693074\\
0.399999999999999	0.933635640396236\\
0.5	0.934956562061311\\
0.6	0.936534216724605\\
0.7	0.938347894757354\\
0.8	0.940374302793025\\
0.899999999999999	0.942588080630007\\
1	0.944962346263369\\
1.1	0.947469252001261\\
1.2	0.950080535137753\\
1.3	0.952768047815287\\
1.4	0.955504252416134\\
1.5	0.958262670951301\\
1.6	0.961018279332583\\
1.7	0.963747839981462\\
1.8	0.966430168815761\\
1.9	0.969046335143017\\
2	0.971579795278059\\
2.1	0.974016462714284\\
2.2	0.976344719359029\\
2.3	0.978555373663212\\
2.4	0.980641572425167\\
2.5	0.982598673639331\\
2.6	0.984424088017149\\
2.7	0.986117096769121\\
2.8	0.987678652946618\\
2.9	0.989111173150996\\
3	0.990418325773942\\
3.1	0.991604821185585\\
3.2	0.99267620847969\\
3.3	0.993638682556733\\
3.4	0.994498904508609\\
3.5	0.995263837489746\\
3.6	0.995940599537484\\
3.7	0.996536334154979\\
3.8	0.997058098900334\\
3.9	0.997512771740884\\
4	0.997906974531462\\
4.1	0.998247012658323\\
4.2	0.998538829650163\\
4.3	0.998787975388986\\
4.4	0.998999586448062\\
4.5	0.999178377034739\\
4.6	0.999328639013572\\
4.7	0.999454249521935\\
4.8	0.999558684759021\\
4.9	0.999645038622378\\
5	0.999716044976099\\
};

\addplot [color=mycolor5, thick]
  table[row sep=crcr]{%
-5	0.999710621843096\\
-4.9	0.999637531814993\\
-4.8	0.9995483804104\\
-4.7	0.999440220971758\\
-4.6	0.999309693930937\\
-4.5	0.999152994561322\\
-4.4	0.998965843338892\\
-4.3	0.998743460252967\\
-4.2	0.998480544595768\\
-4.1	0.99817126194298\\
-4	0.997809240207078\\
-3.9	0.997387576791724\\
-3.8	0.996898858987796\\
-3.7	0.996335199816753\\
-3.6	0.99568829153084\\
-3.5	0.99494947890678\\
-3.4	0.994109854304189\\
-3.3	0.993160376186085\\
-3.2	0.992092012401872\\
-3.1	0.990895909000113\\
-3	0.989563584660279\\
-2.9	0.988087150004965\\
-2.8	0.986459550079403\\
-2.7	0.984674827173979\\
-2.6	0.982728399938732\\
-2.5	0.980617353428289\\
-2.4	0.97834073336561\\
-2.3	0.975899836580005\\
-2.2	0.973298488327767\\
-2.1	0.970543296121529\\
-2	0.967643868863644\\
-1.9	0.964612989589662\\
-1.8	0.96146673006846\\
-1.7	0.958224495955806\\
-1.6	0.954908992222346\\
-1.5	0.951546100216413\\
-1.4	0.948164659987376\\
-1.3	0.944796154360238\\
-1.2	0.941474294649194\\
-1.1	0.938234511715911\\
-1	0.935113360163912\\
-0.899999999999999	0.932147847622879\\
-0.8	0.929374705096509\\
-0.7	0.926829617988735\\
-0.6	0.92454644044886\\
-0.5	0.922556417865304\\
-0.399999999999999	0.92088744350376\\
-0.3	0.919563375292271\\
-0.199999999999999	0.918603437531921\\
-0.0999999999999996	0.918021729860996\\
0	0.917826862204723\\
0.0999999999999996	0.918021729860996\\
0.199999999999999	0.918603437531921\\
0.3	0.919563375292271\\
0.399999999999999	0.92088744350376\\
0.5	0.922556417865305\\
0.6	0.92454644044886\\
0.7	0.926829617988735\\
0.8	0.929374705096507\\
0.899999999999999	0.932147847622879\\
1	0.935113360163911\\
1.1	0.938234511715911\\
1.2	0.941474294649194\\
1.3	0.944796154360237\\
1.4	0.948164659987376\\
1.5	0.951546100216414\\
1.6	0.954908992222345\\
1.7	0.958224495955807\\
1.8	0.961466730068459\\
1.9	0.96461298958966\\
2	0.967643868863644\\
2.1	0.970543296121533\\
2.2	0.973298488327771\\
2.3	0.975899836580009\\
2.4	0.978340733365605\\
2.5	0.980617353428293\\
2.6	0.982728399938735\\
2.7	0.98467482717398\\
2.8	0.986459550079404\\
2.9	0.988087150004966\\
3	0.989563584660281\\
3.1	0.990895909000116\\
3.2	0.992092012401868\\
3.3	0.993160376186085\\
3.4	0.994109854304186\\
3.5	0.994949478906775\\
3.6	0.995688291530836\\
3.7	0.996335199816756\\
3.8	0.996898858987791\\
3.9	0.997387576791718\\
4	0.997809240207076\\
4.1	0.998171261942987\\
4.2	0.998480544595768\\
4.3	0.998743460252961\\
4.4	0.998965843338897\\
4.5	0.999152994561324\\
4.6	0.999309693930932\\
4.7	0.999440220971764\\
4.8	0.999548380410396\\
4.9	0.999637531814991\\
5	0.999710621843103\\
};

\addplot [color=mycolor6, thick]
  table[row sep=crcr]{%
-5	0.999620531033737\\
-4.9	0.999525513986793\\
-4.8	0.999409911810471\\
-4.7	0.99927004121006\\
-4.6	0.999101728715095\\
-4.5	0.99890027886973\\
-4.4	0.998660446987842\\
-4.3	0.99837641810305\\
-4.2	0.998041793907541\\
-4.1	0.99764958961728\\
-4	0.997192242816044\\
-3.9	0.996661636406926\\
-3.8	0.996049137826607\\
-3.7	0.99534565664317\\
-3.6	0.994541722550272\\
-3.5	0.993627585576768\\
-3.4	0.992593340039206\\
-3.3	0.991429073363953\\
-3.2	0.990125040386696\\
-3.1	0.988671863093203\\
-3	0.987060754993869\\
-2.9	0.98528376842828\\
-2.8	0.983334062083888\\
-2.7	0.98120618490152\\
-2.6	0.978896371355636\\
-2.5	0.97640284187432\\
-2.4	0.973726100948445\\
-2.3	0.970869224326991\\
-2.2	0.967838125670657\\
-2.1	0.96464179221074\\
-2	0.961292478411084\\
-1.9	0.957805846434796\\
-1.8	0.954201042447042\\
-1.7	0.950500698503426\\
-1.6	0.946730851026935\\
-1.5	0.942920768689361\\
-1.4	0.939102684881484\\
-1.3	0.935311432842699\\
-1.2	0.931583984851372\\
-1.1	0.927958900540817\\
-1	0.924475693255573\\
-0.899999999999999	0.921174127221164\\
-0.8	0.918093461967995\\
-0.7	0.915271663715242\\
-0.6	0.912744606074778\\
-0.5	0.910545284286195\\
-0.399999999999999	0.908703068082901\\
-0.3	0.907243018102913\\
-0.199999999999999	0.906185289441744\\
-0.0999999999999996	0.905544643510459\\
0	0.90533008588991\\
0.0999999999999996	0.905544643510459\\
0.199999999999999	0.906185289441745\\
0.3	0.907243018102913\\
0.399999999999999	0.908703068082901\\
0.5	0.910545284286195\\
0.6	0.912744606074777\\
0.7	0.915271663715242\\
0.8	0.918093461967993\\
0.899999999999999	0.921174127221164\\
1	0.924475693255572\\
1.1	0.927958900540817\\
1.2	0.931583984851372\\
1.3	0.935311432842697\\
1.4	0.939102684881483\\
1.5	0.942920768689362\\
1.6	0.946730851026934\\
1.7	0.950500698503428\\
1.8	0.954201042447041\\
1.9	0.957805846434794\\
2	0.961292478411084\\
2.1	0.964641792210745\\
2.2	0.967838125670661\\
2.3	0.970869224326995\\
2.4	0.97372610094844\\
2.5	0.976402841874325\\
2.6	0.97889637135564\\
2.7	0.981206184901521\\
2.8	0.983334062083889\\
2.9	0.985283768428281\\
3	0.987060754993872\\
3.1	0.988671863093206\\
3.2	0.99012504038669\\
3.3	0.991429073363953\\
3.4	0.992593340039203\\
3.5	0.993627585576761\\
3.6	0.994541722550268\\
3.7	0.995345656643173\\
3.8	0.996049137826601\\
3.9	0.996661636406919\\
4	0.997192242816042\\
4.1	0.997649589617289\\
4.2	0.998041793907541\\
4.3	0.998376418103042\\
4.4	0.998660446987848\\
4.5	0.998900278869734\\
4.6	0.999101728715089\\
4.7	0.999270041210068\\
4.8	0.999409911810466\\
4.9	0.99952551398679\\
5	0.999620531033746\\
};

\addplot [color=mycolor7, thick]
  table[row sep=crcr]{%
-5	0.999471490528483\\
-4.9	0.999341667507153\\
-4.8	0.999184448609732\\
-4.7	0.998995118922978\\
-4.6	0.998768377974612\\
-4.5	0.998498310350266\\
-4.4	0.998178363309043\\
-4.3	0.997801333302293\\
-4.2	0.997359363428831\\
-4.1	0.996843953955608\\
-4	0.996245988086667\\
-3.9	0.995555775165384\\
-3.8	0.994763113435105\\
-3.7	0.99385737435087\\
-3.6	0.99282761021933\\
-3.5	0.991662686635234\\
-3.4	0.990351440771732\\
-3.3	0.988882866060913\\
-3.2	0.987246323165392\\
-3.1	0.985431776389908\\
-3	0.983430053816504\\
-2.9	0.981233128476434\\
-2.8	0.978834416811223\\
-2.7	0.976229089546763\\
-2.6	0.973414388938626\\
-2.5	0.970389945183303\\
-2.4	0.967158083676886\\
-2.3	0.963724113796004\\
-2.2	0.960096589038542\\
-2.1	0.956287527761354\\
-2	0.952312583458059\\
-1.9	0.948191153599267\\
-1.8	0.943946416570338\\
-1.7	0.939605287236657\\
-1.6	0.935198283174574\\
-1.5	0.93075929563603\\
-1.4	0.926325261848259\\
-1.3	0.92193573823682\\
-1.2	0.917632377518816\\
-1.1	0.913458316228144\\
-1	0.90945748296139\\
-0.899999999999999	0.905673841301405\\
-0.8	0.902150584799654\\
-0.7	0.898929304384651\\
-0.6	0.896049150925415\\
-0.5	0.893546017246866\\
-0.399999999999999	0.89145176453226\\
-0.3	0.889793517663092\\
-0.199999999999999	0.888593052599786\\
-0.0999999999999996	0.887866296417431\\
0	0.887622957161234\\
0.0999999999999996	0.887866296417431\\
0.199999999999999	0.888593052599786\\
0.3	0.889793517663091\\
0.399999999999999	0.89145176453226\\
0.5	0.893546017246867\\
0.6	0.896049150925415\\
0.7	0.898929304384651\\
0.8	0.902150584799652\\
0.899999999999999	0.905673841301405\\
1	0.909457482961388\\
1.1	0.913458316228145\\
1.2	0.917632377518816\\
1.3	0.921935738236819\\
1.4	0.926325261848258\\
1.5	0.930759295636032\\
1.6	0.935198283174572\\
1.7	0.93960528723666\\
1.8	0.943946416570336\\
1.9	0.948191153599265\\
2	0.952312583458059\\
2.1	0.95628752776136\\
2.2	0.960096589038547\\
2.3	0.963724113796009\\
2.4	0.96715808367688\\
2.5	0.970389945183309\\
2.6	0.97341438893863\\
2.7	0.976229089546764\\
2.8	0.978834416811224\\
2.9	0.981233128476435\\
3	0.983430053816508\\
3.1	0.985431776389911\\
3.2	0.987246323165386\\
3.3	0.988882866060912\\
3.4	0.990351440771728\\
3.5	0.991662686635225\\
3.6	0.992827610219325\\
3.7	0.993857374350874\\
3.8	0.994763113435098\\
3.9	0.995555775165374\\
4	0.996245988086665\\
4.1	0.99684395395562\\
4.2	0.997359363428831\\
4.3	0.997801333302282\\
4.4	0.99817836330905\\
4.5	0.99849831035027\\
4.6	0.998768377974603\\
4.7	0.998995118922988\\
4.8	0.999184448609726\\
4.9	0.99934166750715\\
5	0.999471490528496\\
};

\end{axis}
\end{tikzpicture}%

%% file: figs/gts-region.tex
\definecolor{ttzzqq}{rgb}{0.2,0.6,0.0}
\definecolor{qqzzff}{rgb}{0,0.6,1.0}
\definecolor{ccqqqq}{rgb}{0.89,0.0,0.13}

\scalebox{0.825}{
\begin{tikzpicture}[line cap=round,line join=round,x=1.0cm,y=1.0cm]
\begin{axis}[
x=1.0cm,
y=1.0cm,
axis lines=middle,
ymajorgrids=true,
xmajorgrids=true,
xmin=-4.7,
xmax=3.2,
ymin=-3.0,
ymax=3.7,
xtick={-4.0,-3.0,...,3.0},
ytick={-3.0,-2.0,...,3.0},]
\clip(-4.7,-3.) rectangle (3.2,3.7);
\fill[line width=2.pt,color=ttzzqq,fill=ttzzqq,fill opacity=0.10000000149011612] (0.,0.) -- (1.510752824127976,0.8696840754443048) -- (1.4957950162404132,2.5898319825140077) -- (0.,1.722279125035375) -- cycle;
\fill[line width=2.pt,color=ttzzqq,fill=ttzzqq,fill opacity=0.10000000149011612] (0.,0.) -- (-2.497939689738808,-0.8654216395129608) -- (-4.5022859466722,-2.6005273544702265) -- (-2.004332029449241,-1.7329744969915937) -- cycle;
\draw[line width=4.pt,fill=black,fill opacity=0.10000000149011612] (-12.667096430311195,4.229459508474687) -- (-9.822838333224496,4.229459508474687);
\draw[line width=4.pt,fill=black,fill opacity=0.10000000149011612] (-12.524883525456861,3.888148536824283) -- (-9.680625428370162,3.888148536824283);
\draw [->,line width=2.pt] (0.,0.) -- (0.,1.722279125035375);
\draw [->,line width=2.pt] (0.,0.) -- (1.510752824127976,0.8696840754443048);
\draw [->,line width=2.pt] (0.,0.) -- (-2.497939689738808,-0.8654216395129608);
\draw [->,line width=2.pt] (0.,0.) -- (-2.004332029449241,-1.7329744969915937);
\draw [line width=2.pt,color=ttzzqq] (1.510752824127976,0.8696840754443048)-- (1.4957950162404132,2.5898319825140077);
\draw [line width=2.pt,color=ttzzqq] (1.4957950162404132,2.5898319825140077)-- (0.,1.722279125035375);
\draw [line width=2.pt,color=ttzzqq] (-2.497939689738808,-0.8654216395129608)-- (-4.5022859466722,-2.6005273544702265);
\draw [line width=2.pt,color=ttzzqq] (-4.5022859466722,-2.6005273544702265)-- (-2.004332029449241,-1.7329744969915937);
\begin{scriptsize}
\draw [fill=black] (-12.667096430311195,4.229459508474687) circle (2.5pt);
\draw[color=black] (-12.567547396913161,4.5636598348823725) node {$a = -8$};
\draw [fill=black] (-9.680625428370162,3.888148536824283) circle (2.5pt);
\draw[color=black] (-9.623740266428428,4.22234886323197) node {$n = 5$};
\draw [fill=ccqqqq] (-5.,0.) circle (3.0pt);
\draw [fill=ccqqqq] (-4.,0.) circle (3.0pt);
\draw [fill=ccqqqq] (-3.,0.) circle (3.0pt);
\draw [fill=ccqqqq] (-2.,0.) circle (3.0pt);
\draw [fill=ccqqqq] (-1.,0.) circle (3.0pt);
\draw [fill=ccqqqq] (0.,0.) circle (3.0pt);
\draw [fill=ccqqqq] (1.,0.) circle (3.0pt);
\draw [fill=ccqqqq] (2.,0.) circle (3.0pt);
\draw [fill=ccqqqq] (3.,0.) circle (3.0pt);
\draw [fill=ccqqqq] (4.,0.) circle (3.0pt);
\draw [fill=ccqqqq] (4.5023144016405015,0.8547262675567422) circle (3.0pt);
\draw [fill=ccqqqq] (3.5150990810613676,0.8696840754443048) circle (3.0pt);
\draw [fill=ccqqqq] (2.5129259525946717,0.8696840754443048) circle (3.0pt);
\draw [fill=ccqqqq] (1.510752824127976,0.8696840754443048) circle (3.0pt);
\draw [fill=ccqqqq] (0.5085796956612799,0.8547262675567422) circle (3.0pt);
\draw [fill=ccqqqq] (-0.493593432805416,0.8696840754443048) circle (3.0pt);
\draw [fill=ccqqqq] (-1.4957665612721118,0.8696840754443048) circle (3.0pt);
\draw [fill=ccqqqq] (-2.497939689738808,0.8547262675567422) circle (3.0pt);
\draw [fill=ccqqqq] (-3.500112818205504,0.8547262675567422) circle (3.0pt);
\draw [fill=ccqqqq] (-4.5022859466722,0.8696840754443048) circle (3.0pt);
\draw [fill=ccqqqq] (-5.504459075138896,0.8696840754443048) circle (3.0pt);
\draw [fill=ccqqqq] (-4.995893606961767,1.722279125035375) circle (3.0pt);
\draw [fill=ccqqqq] (-3.9937204784950704,1.7372369329229376) circle (3.0pt);
\draw [fill=ccqqqq] (-2.9915473500283745,1.7073213171478123) circle (3.0pt);
\draw [fill=ccqqqq] (-2.004332029449241,1.7372369329229376) circle (3.0pt);
\draw [fill=ccqqqq] (-0.97224328520742,1.722279125035375) circle (3.0pt);
\draw [fill=ccqqqq] (0.,1.722279125035375) circle (3.0pt);
\draw [fill=ccqqqq] (1.0021873559508465,1.7521947408105003) circle (3.0pt);
\draw [fill=ccqqqq] (1.98940267652998,1.722279125035375) circle (3.0pt);
\draw [fill=ccqqqq] (3.0065336128842386,1.722279125035375) circle (3.0pt);
\draw [fill=ccqqqq] (3.9937489334633716,1.7521947408105003) circle (3.0pt);
\draw [fill=ccqqqq] (4.5023144016405015,2.5898319825140077) circle (3.0pt);
\draw [fill=ccqqqq] (3.500141273173805,2.60478979040157) circle (3.0pt);
\draw [fill=ccqqqq] (2.497968144707109,2.60478979040157) circle (3.0pt);
\draw [fill=ccqqqq] (1.4957950162404132,2.5898319825140077) circle (3.0pt);
\draw [fill=ccqqqq] (0.4936218877737173,2.5898319825140077) circle (3.0pt);
\draw [fill=ccqqqq] (-0.493593432805416,2.5898319825140077) circle (3.0pt);
\draw [fill=ccqqqq] (-1.4957665612721118,2.5898319825140077) circle (3.0pt);
\draw [fill=ccqqqq] (-2.5128974976263705,2.60478979040157) circle (3.0pt);
\draw [fill=ccqqqq] (-3.485155010317941,2.619747598289133) circle (3.0pt);
\draw [fill=ccqqqq] (-4.5022859466722,2.60478979040157) circle (3.0pt);
\draw [fill=ccqqqq] (-3.9937204784950704,3.487300455767766) circle (3.0pt);
\draw [fill=ccqqqq] (-3.006505157915937,3.4573848399926406) circle (3.0pt);
\draw [fill=ccqqqq] (-1.9893742215616785,3.4573848399926406) circle (3.0pt);
\draw [fill=ccqqqq] (-0.9872010930949826,3.442427032105078) circle (3.0pt);
\draw [fill=ccqqqq] (0.,3.4573848399926406) circle (3.0pt);
\draw [fill=ccqqqq] (1.0171451638384092,3.4573848399926406) circle (3.0pt);
\draw [fill=ccqqqq] (1.98940267652998,3.4573848399926406) circle (3.0pt);
\draw [fill=ccqqqq] (3.0065336128842386,3.4573848399926406) circle (3.0pt);
\draw [fill=ccqqqq] (4.0087067413509345,3.472342647880203) circle (3.0pt);
\draw [fill=ccqqqq] (4.5023144016405015,4.339895505358836) circle (3.0pt);
\draw [fill=ccqqqq] (3.5150990810613676,4.339895505358836) circle (3.0pt);
\draw [fill=ccqqqq] (2.497968144707109,4.339895505358836) circle (3.0pt);
\draw [fill=ccqqqq] (1.510752824127976,4.309979889583711) circle (3.0pt);
\draw [fill=ccqqqq] (0.4936218877737173,4.324937697471273) circle (3.0pt);
\draw [fill=ccqqqq] (-0.493593432805416,4.339895505358836) circle (3.0pt);
\draw [fill=ccqqqq] (-1.4957665612721118,4.339895505358836) circle (3.0pt);
\draw [fill=ccqqqq] (-2.497939689738808,4.339895505358836) circle (3.0pt);
\draw [fill=ccqqqq] (-3.500112818205504,4.324937697471273) circle (3.0pt);
\draw [fill=ccqqqq] (-4.487328138784637,4.324937697471273) circle (3.0pt);
\draw [fill=ccqqqq] (-4.965977991186641,3.487300455767766) circle (3.0pt);
\draw [fill=ccqqqq] (-5.489501267251333,-0.8654216395129608) circle (3.0pt);
\draw [fill=ccqqqq] (-4.5022859466722,-0.8654216395129608) circle (3.0pt);
\draw [fill=ccqqqq] (-3.500112818205504,-0.8654216395129608) circle (3.0pt);
\draw [fill=ccqqqq] (-2.497939689738808,-0.8654216395129608) circle (3.0pt);
\draw [fill=ccqqqq] (-1.4957665612721118,-0.8654216395129608) circle (3.0pt);
\draw [fill=ccqqqq] (-0.493593432805416,-0.8654216395129608) circle (3.0pt);
\draw [fill=ccqqqq] (0.5085796956612799,-0.8654216395129608) circle (3.0pt);
\draw [fill=ccqqqq] (1.510752824127976,-0.8654216395129608) circle (3.0pt);
\draw [fill=ccqqqq] (2.5129259525946717,-0.8654216395129608) circle (3.0pt);
\draw [fill=ccqqqq] (3.5150990810613676,-0.8803794474005234) circle (3.0pt);
\draw [fill=ccqqqq] (4.5023144016405015,-0.8654216395129608) circle (3.0pt);
\draw [fill=ccqqqq] (-5.998066735428463,-1.7329744969915937) circle (3.0pt);
\draw [fill=ccqqqq] (-4.995893606961767,-1.7329744969915937) circle (3.0pt);
\draw [fill=ccqqqq] (-3.9937204784950704,-1.7329744969915937) circle (3.0pt);
\draw [fill=ccqqqq] (-2.9915473500283745,-1.7329744969915937) circle (3.0pt);
\draw [fill=ccqqqq] (-2.004332029449241,-1.7329744969915937) circle (3.0pt);
\draw [fill=ccqqqq] (-1.0021589009825453,-1.7329744969915937) circle (3.0pt);
\draw [fill=ccqqqq] (0.,-1.7329744969915937) circle (3.0pt);
\draw [fill=ccqqqq] (1.0021873559508465,-1.7329744969915937) circle (3.0pt);
\draw [fill=ccqqqq] (2.019318292305105,-1.718016689104031) circle (3.0pt);
\draw [fill=ccqqqq] (3.0065336128842386,-1.7329744969915937) circle (3.0pt);
\draw [fill=ccqqqq] (3.9937489334633716,-1.7329744969915937) circle (3.0pt);
\draw [fill=ccqqqq] (-5.489501267251333,-2.6005273544702265) circle (3.0pt);
\draw [fill=ccqqqq] (-4.5022859466722,-2.6005273544702265) circle (3.0pt);
\draw [fill=ccqqqq] (-3.500112818205504,-2.6005273544702265) circle (3.0pt);
\draw [fill=ccqqqq] (-2.497939689738808,-2.615485162357789) circle (3.0pt);
\draw [fill=ccqqqq] (-1.4957665612721118,-2.6005273544702265) circle (3.0pt);
\draw [fill=ccqqqq] (-0.493593432805416,-2.6005273544702265) circle (3.0pt);
\draw [fill=ccqqqq] (0.5085796956612799,-2.6005273544702265) circle (3.0pt);
\draw [fill=ccqqqq] (1.4957950162404132,-2.6005273544702265) circle (3.0pt);
\draw [fill=ccqqqq] (2.5129259525946717,-2.6005273544702265) circle (3.0pt);
\draw [fill=ccqqqq] (3.500141273173805,-2.6005273544702265) circle (3.0pt);
\draw [fill=ccqqqq] (4.5023144016405015,-2.6005273544702265) circle (3.0pt);
\draw [fill=ccqqqq] (-5.998066735428463,-3.4680802119488594) circle (3.0pt);
\draw [fill=ccqqqq] (-4.995893606961767,-3.4680802119488594) circle (3.0pt);
\draw [fill=ccqqqq] (-3.9937204784950704,-3.4680802119488594) circle (3.0pt);
\draw [fill=ccqqqq] (-2.9915473500283745,-3.4680802119488594) circle (3.0pt);
\draw [fill=ccqqqq] (-2.004332029449241,-3.4680802119488594) circle (3.0pt);
\draw [fill=ccqqqq] (-0.9872010930949826,-3.4680802119488594) circle (3.0pt);
\draw [fill=ccqqqq] (1.4227484150666597E-5,-3.4680802119488594) circle (3.0pt);
\draw [fill=ccqqqq] (1.0021873559508465,-3.4680802119488594) circle (3.0pt);
\draw [fill=ccqqqq] (1.98940267652998,-3.4680802119488594) circle (3.0pt);
\draw [fill=ccqqqq] (3.0065336128842386,-3.4680802119488594) circle (3.0pt);
\draw [fill=ccqqqq] (4.0087067413509345,-3.4531224040612964) circle (3.0pt);
\draw [fill=ccqqqq] (-5.504459075138896,2.60478979040157) circle (3.0pt);
\draw [fill=ccqqqq] (-5.489501267251333,4.339895505358836) circle (3.0pt);

grid = both, grid style={dotted,black!50},
every axis x label/.style={at={(current axis.south east)},right=1mm},
every axis y label/.style={at={(current axis.north west)},above=1mm},
enlargelimits={abs=30pt,upper},
]
\addplot [color=qqzzff, ultra thick, only marks,mark=x,mark size = 4pt] coordinates {
    (-1,0) (1,0)  (0.5160398496856559,0.8590136634269516) (0.5160398496856559,-0.8617624853104994) (-0.49367177478012236,-0.8617624853104994) (-0.49367177478012236,0.873234953912385)
};
\end{scriptsize}
\end{axis}
\end{tikzpicture}}

%% file: figs/bad_secrecy_gain_bordered_double_circulant_n6.tex
%
%
\begin{tikzpicture}
\begin{axis}[%
width=8.00cm,
height=5.00cm,
tick align=outside,
tick pos=left,
xmin=-5,
xmax=5,
xlabel style={font=\color{white!15!black}},
xlabel={$\tau$ (dB)},
ymin=1,
ymax=1.003,
ytick={1.0, 1.001, 1.002, 1.003},        
yticklabels={1.0, 1.001, 1.002, 1.003},   
y tick label style={font=\scriptsize, xshift=5pt},  
ylabel style={font=\color{white!15!black}, yshift=-0.10cm},
ylabel={Theta series ratio},
axis background/.style={fill=white},
xmajorgrids,
ymajorgrids,
legend style={at={(0.434,1.00)}, legend cell align=left, align=left, draw=white!15!black}
]
\addplot [color=blue, mark=*, mark size=1pt, mark options={solid, fill=blue, blue}]
  table[row sep=crcr]{%
-5	1.00000033482678\\
-4.9	1.00000048799349\\
-4.8	1.00000070422967\\
-4.7	1.00000100646516\\
-4.6	1.00000142476281\\
-4.5	1.00000199810633\\
-4.4	1.00000277649682\\
-4.3	1.00000382336924\\
-4.2	1.00000521832628\\
-4.1	1.00000706016956\\
-4	1.00000947018473\\
-3.9	1.00001259561\\
-3.8	1.0000166131854\\
-3.7	1.00002173264459\\
-3.6	1.00002819997305\\
-3.5	1.00003630021735\\
-3.4	1.0000463595931\\
-3.3	1.00005874660593\\
-3.2	1.00007387187433\\
-3.1	1.00009218632872\\
-3	1.00011417746159\\
-2.9	1.00014036332274\\
-2.8	1.00017128399494\\
-2.7	1.00020749035108\\
-2.6	1.0002495299868\\
-2.5	1.00029793034136\\
-2.4	1.00035317916414\\
-2.3	1.00041570264968\\
-2.2	1.00048584174529\\
-2.1	1.00056382732317\\
-2	1.00064975509347\\
-1.9	1.00074356130211\\
-1.8	1.00084500039502\\
-1.7	1.00095362592308\\
-1.6	1.00106877599623\\
-1.5	1.00118956455917\\
-1.4	1.00131487964509\\
-1.3	1.00144338956514\\
-1.2	1.00157355770876\\
-1.1	1.00170366627355\\
-1	1.00183184882441\\
-0.899999999999999	1.00195613112337\\
-0.8	1.00207447919742\\
-0.7	1.00218485315366\\
-0.6	1.00228526484033\\
-0.5	1.00237383712044\\
-0.399999999999999	1.00244886230032\\
-0.3	1.00250885716098\\
-0.199999999999999	1.00255261208955\\
-0.0999999999999996	1.00257923200573\\
0	1.00258816711777\\
0.0999999999999996	1.00257923200573\\
0.199999999999999	1.00255261208955\\
0.3	1.00250885716098\\
0.399999999999999	1.00244886230032\\
0.5	1.00237383712044\\
0.6	1.00228526484033\\
0.7	1.00218485315366\\
0.8	1.00207447919742\\
0.899999999999999	1.00195613112337\\
1	1.00183184882441\\
1.1	1.00170366627355\\
1.2	1.00157355770876\\
1.3	1.00144338956514\\
1.4	1.00131487964509\\
1.5	1.00118956455917\\
1.6	1.00106877599624\\
1.7	1.00095362592308\\
1.8	1.00084500039502\\
1.9	1.00074356130211\\
2	1.00064975509347\\
2.1	1.00056382732317\\
2.2	1.00048584174529\\
2.3	1.00041570264968\\
2.4	1.00035317916414\\
2.5	1.00029793034136\\
2.6	1.0002495299868\\
2.7	1.00020749035108\\
2.8	1.00017128399494\\
2.9	1.00014036332274\\
3	1.00011417746159\\
3.1	1.00009218632872\\
3.2	1.00007387187433\\
3.3	1.00005874660593\\
3.4	1.0000463595931\\
3.5	1.00003630021735\\
3.6	1.00002819997305\\
3.7	1.00002173264459\\
3.8	1.0000166131854\\
3.9	1.00001259561\\
4	1.00000947018473\\
4.1	1.00000706016956\\
4.2	1.00000521832628\\
4.3	1.00000382336924\\
4.4	1.00000277649682\\
4.5	1.00000199810633\\
4.6	1.00000142476281\\
4.7	1.00000100646516\\
4.8	1.00000070422967\\
4.9	1.00000048799349\\
5	1.00000033482678\\
};
\addlegendentry{$\Delta_{\ConstrAq[4]{\code{C}_3}}(\tau)$}

\end{axis}
\end{tikzpicture}%

%% file: figs/conj1_secrecy_gain_bordered_double_circulant_n6.tex
%
%
\definecolor{mycolor1}{rgb}{0.00000,0.44700,0.74100}%
\definecolor{mycolor2}{rgb}{0.85000,0.32500,0.09800}%
\definecolor{mycolor3}{rgb}{0.92900,0.69400,0.12500}%
\definecolor{mycolor4}{rgb}{0.49400,0.18400,0.55600}%
\definecolor{mycolor5}{rgb}{0.46600,0.67400,0.18800}%
\definecolor{mycolor6}{rgb}{0.30100,0.74500,0.93300}%
\definecolor{mycolor7}{rgb}{0.63500,0.07800,0.18400}%
\begin{tikzpicture}[spy using outlines={circle, magnification=3.5, size=1.25cm, connect spies}]

\begin{axis}[%
width=8cm,
height=5.00cm,
tick align=outside,
tick pos=left,
xmin=-5,
xmax=5,
xlabel style={font=\color{white!15!black}},
xlabel={$\tau$ (dB)},
ymin=0.88,
ymax=1.02,
y tick label style={font=\footnotesize, xshift=5pt},  
ylabel style={font=\color{white!15!black}, yshift=-0.10cm},
ylabel={Theta series ratio},
xmajorgrids,
ymajorgrids,
legend style={at={(0.435,0.178)}, legend cell align=left, align=left, draw=white!15!black}
]
\addplot [color=mycolor1, thick]
  table[row sep=crcr]{%
-5	1.00063226663784\\
-4.9	1.00071669207712\\
-4.8	1.00080691031914\\
-4.7	1.00090222395779\\
-4.6	1.00100166025053\\
-4.5	1.00110393981534\\
-4.4	1.00120744755188\\
-4.3	1.00131020695869\\
-4.2	1.00140985910285\\
-4.1	1.00150364755695\\
-4	1.00158841064613\\
-3.9	1.00166058233945\\
-3.8	1.00171620307064\\
-3.7	1.00175094167795\\
-3.6	1.0017601295091\\
-3.5	1.00173880754045\\
-3.4	1.00168178710965\\
-3.3	1.00158372455636\\
-3.2	1.00143920970827\\
-3.1	1.00124286774317\\
-3	1.00098947350732\\
-2.9	1.00067407688469\\
-2.8	1.00029213730184\\
-2.7	0.999839664933157\\
-2.6	0.999313365658214\\
-2.5	0.998710786336408\\
-2.4	0.998030456526683\\
-2.3	0.997272022415203\\
-2.2	0.996436368446748\\
-2.1	0.995525722010882\\
-2	0.994543736534832\\
-1.9	0.99349554850177\\
-1.8	0.992387804261462\\
-1.7	0.991228653039322\\
-1.6	0.990027703281356\\
-1.5	0.988795940388648\\
-1.4	0.987545604978272\\
-1.3	0.986290032029614\\
-1.2	0.985043452597468\\
-1.1	0.983820761148101\\
-1	0.982637252945922\\
-0.899999999999999	0.981508337225325\\
-0.8	0.980449233061133\\
-0.7	0.979474655839186\\
-0.6	0.978598502968484\\
-0.5	0.977833547918323\\
-0.399999999999999	0.97719115177154\\
-0.3	0.976681001235824\\
-0.199999999999999	0.976310881444696\\
-0.0999999999999996	0.976086490921881\\
0	0.976011304809553\\
0.0999999999999996	0.97608649092188\\
0.199999999999999	0.976310881444696\\
0.3	0.976681001235824\\
0.399999999999999	0.97719115177154\\
0.5	0.977833547918323\\
0.6	0.978598502968484\\
0.7	0.979474655839187\\
0.8	0.980449233061132\\
0.899999999999999	0.981508337225326\\
1	0.982637252945922\\
1.1	0.983820761148101\\
1.2	0.985043452597468\\
1.3	0.986290032029614\\
1.4	0.987545604978272\\
1.5	0.988795940388648\\
1.6	0.990027703281355\\
1.7	0.991228653039322\\
1.8	0.992387804261462\\
1.9	0.993495548501769\\
2	0.994543736534832\\
2.1	0.995525722010883\\
2.2	0.996436368446749\\
2.3	0.997272022415204\\
2.4	0.998030456526681\\
2.5	0.998710786336409\\
2.6	0.999313365658215\\
2.7	0.999839664933157\\
2.8	1.00029213730184\\
2.9	1.00067407688469\\
3	1.00098947350732\\
3.1	1.00124286774317\\
3.2	1.00143920970827\\
3.3	1.00158372455636\\
3.4	1.00168178710965\\
3.5	1.00173880754044\\
3.6	1.0017601295091\\
3.7	1.00175094167795\\
3.8	1.00171620307064\\
3.9	1.00166058233945\\
4	1.00158841064613\\
4.1	1.00150364755695\\
4.2	1.00140985910285\\
4.3	1.00131020695869\\
4.4	1.00120744755188\\
4.5	1.00110393981534\\
4.6	1.00100166025053\\
4.7	1.00090222395779\\
4.8	1.00080691031914\\
4.9	1.00071669207712\\
5	1.00063226663784\\
};
\addlegendentry{$\Delta_{\ConstrAq[4]{\code{C}_4}}(\tau)$}

\end{axis}
\spy [magenta] on (5.50,3.00) in node [left] at (5.75,1.5); 
\end{tikzpicture}%